\documentclass{ifacconf}

\usepackage[round]{natbib} 
\usepackage{amsmath,amssymb}
\usepackage{graphicx}	
\usepackage[table,xcdraw]{xcolor}	
\usepackage{caption}
\usepackage[skip=0cm,list=true,labelfont=it]{subcaption}
\usepackage{balance}
\usepackage{url}

\begin{document}
\begin{frontmatter}

\title{Analytical Derivation and Comparison of \\ Alarm Similarity Measures\thanksref{licence}} 

\thanks[licence]{Copyright \copyright  \, 2021 The Authors. This work has been accepted to the 16th IFAC Symposium on Advanced Control of Chemical Processes as an open access article under the CC-BY-NC-ND license.}

\author[IUT]{Amir Hossein Kargaran} 
\author[IUT]{Amir Neshastegaran} 
\author[IUT]{Iman Izadi}
\author[IUT]{Ehsan Yazdian}

\address[IUT]{Department of Electrical and Computer Engineering\\
Isfahan University of Technology
Isfahan 84156-83111, Iran \\ (corresponding author e-mail: iman.izadi@iut.ac.ir)}

\begin{abstract}                
An industrial process includes many devices, variables, and sub-processes that are physically or electronically interconnected. These interconnections imply some level of correlation between different process variables. Since most of the alarms in a process plant are defined on process variables, alarms are also correlated. However, this can be a nuisance to operators, for one fault might trigger a, sometimes large, number of alarms. So, it is essential to find and correct correlated alarms. In this paper, we study different methods and techniques proposed to measure correlation or similarity between alarms. The similarity indices are first analytically calculated and then studied and compared. The results are also validated using Monte-Carlo simulation.
\end{abstract}


\end{frontmatter}

\section{Introduction}
An industrial process often contains a large number of devices, variables, and control algorithms. In such a process, it is important to monitor the behavior of the system to avoid damages, shutdowns, production loss, and safety hazards. The most common method of process monitoring is to define a safe or efficient range for each variable,~and if it exceeded the safe range, notify the operator or other plant staff about the anomaly. The alarm system is designed to perform this task \citep{ci20,izadi-intro}. Once an alarm is activated, the operator can take necessary corrective actions to remedy the~ situation. 

In an industrial process, there are many interconnections between different variables, due to material, energy, or information flow. These interconnections impose some correlation between different process variables. Also, thanks to the advanced technologies and modern distributed control systems (DCS), almost all process variables in a plant are constantly measured (in fact, anything that can be measured is measured). And for any variable that is measured, alarms can be defined. These alarms are known as process alarms. If two process variables are connected and correlated, it is expected that their alarms are also correlated, although with different intensity, depending on the corresponding alarm limits. In any case, correlation of alarms are widely reported and observed in industry \citep{Li, Yu, ci7, ci5, yang:2019}. There are other types of alarms in a plant that are not associated with any process variable. They are typically known as system alarms or digital alarms. In this paper, we only focus on process alarms as they count for the majority of alarms in a plant. 

Generally, in a plant, two types of data are generated: process data (the measured values of different process variables), and alarm data (the information of alarms activated in the plant). 

Although more measured variables provide a valuable source of historical information and help operators and engineers to better run their plant, more alarms only distract operators from their normal tasks. An excessive number of alarms, many of which are correlated, is a problem in today's industry.  During alarm rationalization, a process of reviewing and redesigning the alarms in a plant, many of these correlated alarms are eliminated \citep{ISA18.2}. For that purpose, finding correlated alarms is a common task in monitoring alarm systems.

Different methods for correlation analysis in a process are based on these two sets of data. Most of the methods are, obviously, based on process data, because of the large amount of information they carry. However, alarm correlations are also important as they help to understand, maintain, improve or even redesign the alarm system. A few methods to measure the similarity between alarm sequences have been introduced in the literature, including the Pearson correlation of binary alarm sequences \citep{ci4}, Jaccard similarity of binary alarm sequences \citep{ci9,ahmed}, the Pearson correlation of multivalued alarm sequences \citep{ci11}, and the Pearson correlation of continuous-valued alarm sequences \citep{ci7}.

In this paper, we try to answer several questions regarding the similarity of alarms: Can it be verified that if two process variables are correlated, then so are their alarms? If the answer is positive, which measure of similarity does better preserve the correlation? Moreover, what is the role of the alarm limit and how can it affect the alarm similarity. To answer these questions, different methods of correlation analysis of alarm data are mathematically studied. We first review how alarm data is converted to sequences or time-series which is more suitable for mathematical analysis. Then we discuss different indices to measure the similarity between alarm sequences. 
A Monte-Carlo simulation is presented to validate the analytical derivations, and to compare different similarity measures.  

\section{Alarm Data}
An alarm is a notification to the operator about an anomaly in the system. Alarms are almost always a text message displayed to the operator, along with other audio and/or visual signs. An alarm message contains some information about the abnormality, including associated process variable or tag, time of occurrence, alarm or tag description, priority, and value. This information, in addition to being displayed to the operator, is also stored either in flat files or in a relational database. 

Three different entities should be distinguished:
\begin{itemize}
\item \textbf{Process variables (PV)}: In the modern process industry, each process variable (for instance the level of a vessel) is continuously measured by the distributed control system (DCS). The time-stamped measurements are known as process data and are available from databases. The process data is a discrete time-series and shows the trend of each variable over time.

\item \textbf{Individual alarm (IA) tags}: Alarms are often triggered if a PV exceeds its predefined limits. For each PV many different alarms can be defined. For instance, when a PV is higher than a certain limit, a high (HI) alarm is activated. If it continues to rise and passes another higher limit, a high-high (HH) alarm is triggered. Similarly, a PV might have low (LO) or low-low (LL) alarms. Some DCS brands allow up to 16 different alarms to be defined for each PV. Each one of these alarms is referred to as an individual alarm (IA) tag. Therefore, each PV can have multiple IA tags. On average, in a normal size plant, the number of IA tags is about 3-5 times the number of PVs. 

\item \textbf{Collective alarm (CA) tags}: As discussed, multiple IA tags can be associated with one single PV. We can collect all the IA tags that are associated with a PV under one single tag. We refer to this tag as a collective alarm (CA) tag. In this study, we assume that there are only four IA tags for each PV, namely: LL, LO, HI, HH. Therefore, each CA tag for a certain PV contains four IA tags related to that PV.

\end{itemize}

In a plant, correlation analysis and similarity measures can be performed on each set of the aforementioned entities. In this study, we focus on similarity measures between alarm tags (IA or CA) and their relationship to their corresponding PVs. 

\section{Preparation of Alarm Data}
As previously mentioned, due to their nature (a time-stamped message) alarm data (individual or collective tags) cannot be directly used for mathematical analysis. So, we first need to convert this data to a form, often a time-series,  suitable for calculations. For this purpose, a few methods are available in the literature that convert alarm data to: binary \citep{ci5,ci6,ci8,ci9},  multivalued \citep{ci11} or continuous-valued \citep{ci7,Hu-delay} sequences. In this paper, we are interested in the first two methods, which will be reviewed in this section.

Notice that the time scale of the obtained time-series is based on the sampling time of the process and the resolution of alarm time-stamps. 

\subsection{Binary alarm sequence}

For each IA tag, i.e., $IA_i$, we can assign a binary sequence $s_{i}(l)$ of ones and zeros, where $l$ denotes the sampling instance. If the alarm is active for this individual tag at time $l$ then $s_{i}(l)=1$; otherwise $s_{i}(l)=0$. Binary sequences can similarly be defined for CA tags. So, for each CA tag, $CA_i$, the binary sequence $s_{i}(l)$ is defined such that, $s_{i}(l)=1$ if at least one of the IA tags that compose the CA tag is active. Otherwise, i.e., when all the IA tags within the CA tag are inactive, $s_{i}(l)=0$. Therefore, the binary sequence of an alarm  tag (IA or CA) is constructed as
\begin{equation}
 s_{i}(l)=
\begin{cases} 
      1 & \mbox{if alarm is active at $l$}\\
      0 & \mbox{otherwise}\\
\end{cases}
\end{equation}

There are two technical issues here that need to be discussed: point-based vs. interval-based data; and raw vs. padded data. 

\subsubsection{Point-based vs. interval-based data}
An alarm is a message generated at an instant of time. So for each alarm message, there is only one time-stamp. On the other hand, in many DCS brands, once a process variable returns to  normal (within the predefined limits), another message, known as the RTN (return-to-normal) is generated and stored in the database. If the RTN message is available, then we know when the alarm state is over. In this case, alarm duration can be defined from the time of the alarm message to the time of the RTN message. Hence, we have two types of alarm data; point-based (i.e., only the alarm messages are available) and interval-based (both the alarm and RTN messages are available) \citep{man:2019}.

The difference is significant when a binary sequence is constructed. For point-based data, 1's appear only at time instants of the alarm messages. So the 1's are very sparse which makes similarity analysis rather difficult. However, for interval-based data, 1's appear for the alarm duration, and there are many more 1's for each single alarm sequence. Therefore, similarity analysis is more conclusive and meaningful.

\subsubsection{Raw vs. padded data}
If only point-based data is available, or if for some reason it is decided to discard RTN messages, then the binary sequences should be enriched with more 1's so that similarity methods can be used. This is known as padding. For this purpose, a number of 1's are added before and after each alarm sample \citep{ci9}. In this case, for each 1 in the raw alarm sequence, $r$ samples before and after are populated with 1's as well. The selection of $r$ depends on the individual PV and the DCS configuration and requires a better understanding of the process. $r$ is an indication of how long an average alarm might last for a PV. For a fast PV, e.g., pressure or flow, a smaller $r$ is preferred. But for slower variables such as temperature and level, larger $r$ might show better results. An acceptable initial guess could be selected based on the \cite{ISA18.2} recommendations for delay-timers, i.e., 60 seconds for temperature and level variables and 15 seconds for flow and pressure tags, or a fraction of thereof. 

Even if interval-based data is available, padding might still be useful. It helps better analysis by further enrichment of data. Padding is also useful when there are different time lags between different PVs, which is prevalent in processes. 

As an example, consider the two binary alarm sequences:
\begin{center}
 s1: \texttt{01000100100000100}
 
 s2: \texttt{00100010001000010}
\end{center}
By visual inspection, it seems that there is some correlation between the two sequences. But because the alarms in the second sequence are lagged by a sample or two compared to the first one, the calculated correlation might not be high. But after padding, there are more matched 1's and the correlation will be higher. 

\begin{figure}[!t]
\centering
\includegraphics[width=9cm,height=5.5cm]{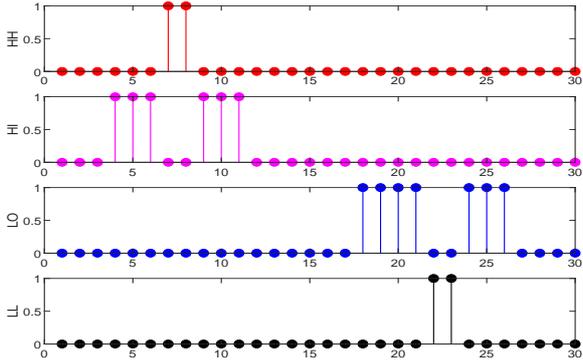}
\vspace{-0.8cm}
\caption{Binary alarm sequences for four IA tags}
\label{HH-HI-LO-LL}
\end{figure}

\subsection{Multivalued alarm sequence}
This method is proposed only for CA tags and is not applicable to IA tags. Each CA tag is a collection of four IA tags, LL, LO, HI, HH, for a PV. So, an alarm could be due to the PV exceeding any of the four limits. Now, we can define a multivalued time sequence as 
\begin{equation}
s_{i}(l)=
\begin{cases}
	2 & \mbox{ if HH alarm is active}\\
	1 & \mbox{ if HI alarm is active}\\
	0 & \mbox{ if no alarm is active}\\
	-1 & \mbox{ if LO alarm is active}\\
	-2 & \mbox{ if LL alarm is active}\\
\end{cases}
\end{equation}
This sequence collects the information of all the alarm tags related to one PV \citep{ci11}. Notice that, it is assumed that at each sampling instant, only one of the alarms can be activated (e.g., once the HH alarm is triggered, the HI alarm will be cleared). This is demonstrated in Fig.~\ref{HH-HI-LO-LL}.

A multivalued sequence, compared to a binary sequence, carries more information regarding the behavior of a PV. A binary sequence indicates only if an alarm is active or not. But a multivalued sequence, in addition to that, shows which individual alarm is active. 

\begin{figure}[!t]
\centering
\includegraphics[width=9cm,height=5.5cm]{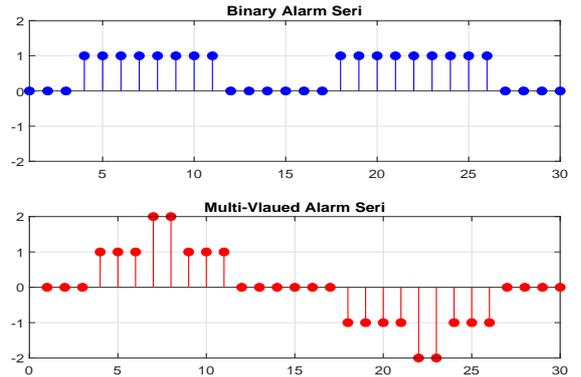}
\vspace{-0.8cm}
\caption{Binary alarm sequence (top) and multivalued alarm sequence  (bottom) for a CA tag}
\label{Alarm-binary-multivalued}
\end{figure}

For example, consider the four IA tags for a PV. The binary sequences for each of these 4 individual alarms are depicted in Fig.~\ref{HH-HI-LO-LL}. If we combine these 4 IA tags into a CA tag, then we can construct the binary sequence of this CA tag as illustrated in  Fig.~\ref{Alarm-binary-multivalued} (top). The multivalued alarm sequence for this CA tag is also shown in Fig.~\ref{Alarm-binary-multivalued} (bottom).

\section{Similarity Measures}
After the alarm sequence  (binary, multivalued) is prepared, the similarity between the two alarm sequences is calculated. Depending on the definition of the sequence, different similarity measures can be used. For binary sequences, binary similarity measures (e.g., the Jaccard index) \citep{ci9,ahmed}, or the Pearson correlation coefficient \citep{ci4} have been used. The latter has been suggested for multivalued alarm sequences \citep{ci11} as well.

Before the methods to calculate similarity are discussed, it should be mentioned that for better similarity analysis a time lag is preferably considered. The reason is that the similarity between two PVs or alarm sequences is not necessarily observed at the same time. A PV might follow the variations of another one with some delay. This is also true for alarm sequences. For instance, if the fuel feed of a furnace that heats a vessel increases, the temperature of the vessel should increase as well. But this happens after some time lag, that should be considered in similarity calculations. However, in the following analytical derivations,  for simplicity, we do not consider time-lags. All the calculations remain valid in case a time-lag between the sequences exists, by considering a shifted sequence.


To calculate the similarity measures of two alarm sequences, we assume that underlying process variables follow a Gaussian distribution. Define ${F}$ and $F_{2d}$ as: 
\begin{equation}
F(x,y) =\begin{cases}
    {\frac {1}{\sqrt {2\pi}}}\int _{x}^{y} 	    e^{(-{\frac {u^{2}}{2}})}du,&  y\geq x\\
    0,              & \text{otherwise}
\end{cases}
\end{equation}

\begin{equation} \begin{split}
    & F_{2d}(x_{a},x_{b},y_{a},y_{b},\rho) = \\
    &  \qquad \qquad \frac{1}{2 \pi \sqrt{1-\rho^2}}
      \int_{x_{a}}^{x_{b}} \int_{y_{a}}^{y_{b}} 
     e^{(-\frac{u^{2} + v^{2}-2\rho u v}{2(1-\rho^2)})} du dv 
\end{split} \end{equation}



Gaussian distribution is assumed for brevity. The process variables can follow any distribution. In this case, ${F}$ and $F_{2d}$ are be defined  accordingly. 
In the following sections, different methods to calculate the similarity between alarm sequences, are discussed and calculated.

\section{Pearson Correlation Coefficient}
The Pearson correlation coefficient is the most common method to calculate the correlation between two variables. It can be used for all alarm sequences as well \citep{ci4}. This applies to all types of binary, multivalued, and continuous-valued alarm sequences.

For two random variables $x$ and $y$, 
the  Pearson correlation coefficient is calculated as 
\begin{equation} \label{Pearson}
\rho_{x, y}=\frac{\mathrm{cov}(x,y)}{\sigma_{x}\sigma_{y}} = \frac{\mathrm{E}[xy]-\mu_{x}\mu_{y}}{\sigma_{x}\sigma_{y}}
\end{equation}
where $\mu_{z}$, and $\sigma_{z}$ are the mean and the standard deviations of $z$, respectively. 


Let $X_{i}\sim N(\mu_{i},\sigma_{i})$ and $X_{j}\sim N(\mu_{j},\sigma_{j})$ be two correlated PVs with Gaussian distribution  and the Pearson correlation coefficient  $\rho$  (i.e., every sample of $X_{i}$ is correlated with its concurrent sample of $X_{j}$ with coefficient $\rho$  and independent form other samples). 

\subsection{Binary alarm sequences} 

Assume that there are two high alarm limits configured for these two PVs with thresholds set at $h_{i}$ and $h_{j}$. Let $B_{i}$ and $B_{j}$ be the two corresponding binary alarm sequences. The question we try to answer here is: what is the Pearson correlation coefficient of $B_{i}$ and $B_{j}$?

It is a well known fact that $B_{i}$ has a Bernoulli distribution, which takes the value 1 with probability $\Pr(X_{i}>h_{i})$ and the value 0 with probability $1-\Pr(X_{i}>h_{i})$. Therefor, the mean and standard deviations of $B_{i}$ can be calculated as 
\begin{equation*}
\begin{split}
\mu_{B_{i}}&=\mathrm{E}[B_{i}]=\Pr(B_{i}=1)\times 1+\Pr(B_{i}=0)\times 0 \\
&=\Pr(X_{i}>h_{i})= F(\frac{h_{i}-\mu_{i}}{\sigma_{i}},+\infty) \\
\sigma^{2}_{B_{i}} &=\mathrm{E} [{B_{i}}^{2}]-\mathrm{E}[B_{i}]^{2}=\Pr(X_{i}>h_{i})-{\Pr(X_{i}>h_{i})}^{2}\\
&=\Pr(X_{i}>h_{i})(1-{\Pr(X_{i}>h_{i})})\\
&=F(\frac{h_{i}-\mu_{i}}{\sigma_{i}},+\infty)(1-F(\frac{h_{i}-\mu_{i}}{\sigma_{i}},+\infty))
\end{split}
\end{equation*}

Therefore, the Pearson correlation coefficient of $B_{i}$ and $B_{j}$ can be calculated as 
\begin{equation*} 
\begin{split}
\rho_{B_{i}, B_{j}}&=\frac{1}{\sigma_{B_{i}}\sigma_{B_{j}}} \biggl[\Pr(X_{i}>h_{i},X_{j}>h_{j})-\mu_{B_{i}}\mu_{B_{j}}\biggr] \\
&= \frac{1}{\sigma_{B_{i}}\sigma_{B_{j}}} \times \\
&\left\{\begin{array}{ll}
F_{2d}(\frac{h_{i}-\mu _{i}}{\sigma _{i}},+\infty,\frac{h_{j}-\mu _{j}}{\sigma _{j}},+\infty,\rho)- \mu_{B_{i}}\mu_{B_{j}} & \quad \rho \neq 1
\\\\
	F(\frac{\mathrm{max}(h_{i},h_{j})-\mu_{i}}{\sigma_{i}},+\infty)-\mu_{B_{i}}\mu_{B_{j}} & \quad \rho = 1
\end{array} \right.
\end{split}
\end{equation*}


\subsection{Multivalued alarm sequence}
Now, assume that alarms, namely high-high, high, low, low-low, are configured for the PVs $X_{i}$ and $X_{j}$. The corresponding threshold are $hh_i$, $h_i$, $l_i$, $ll_i$ for  $X_{i}$, and similarly for  $X_{j}$. With this alarm configuration, two multivalued alarm sequences $M_{i}$ and $M_{j}$ are obtained. It can be easily verified that $M_{i}= 2B_{hh_i} + B_{h_i} - B_{l_i} - 2B_{ll_i}
$
where, for instance $B_{hh_i}$ is the corresponding binary sequence for the HH alarm. The other binary sequences are defined similarly.  The mean $\mu_{M_{i}}$ and  standard deviation $\sigma_{M_{i}}$ of $M_{i}$ can be calculated from 
\begin{equation*}
\begin{split}
\mu_{M_{i}}&=\mathrm{E}[M_{i}]=\sum \limits_{r=-2}^{2} r\, F(S_i^{(r)},S_i^{(r+1)})\\
\sigma^{2}_{M_{i}} &=\mathrm{E}[{M_{i}}^{2}]-\mathrm{E}[M_{i}]^{2} \\
&\!\!=\sum \limits_{r=-2}^{2} r^{2}\, F(S_i^{(r)},S_i^{(r+1)})- \left(\sum \limits_{r=-2}^{2} r\, F(S_i^{(r)},S_i^{(r+1)})\right)^{2}
\end{split}
\end{equation*}
where $S_i^{(r)}$ is defined as 
\begin{equation}
S_i^{(r)}=
\begin{cases}
	+\infty & \mbox{ r=3}\\
	\frac{{hh_{i}-\mu_{i}}}{\sigma_{i}} & \mbox{ r=2}\\
	\frac{{h_{i}-\mu_{i}}}{\sigma_{i}} & \mbox{ r=1}\\
	\frac{{l_{i}-\mu_{i}}}{\sigma_{i}} & \mbox{ r=0}\\
	\frac{{ll_{i}-\mu_{i}}}{\sigma_{i}} & \mbox{ r=-1}\\
	-\infty & \mbox{ r=-2}\\
\end{cases}
\end{equation}


Therefore, the Pearson correlation coefficient of $M_{i}$ and $M_{j}$
can be calculated as
\begin{equation*} \begin{split}
&\rho_{M_{i}, M_{j}}= \frac{1}{\sigma_{M_{i}}\sigma_{M_{j}}}\biggl[\mathrm{E}[M_{i}M_{j}]-\mu_{M_{i}}\mu_{M_{j}}\biggr]=\frac{1}{\sigma_{M_{i}}\sigma_{M_{j}}} \times \\
& ~ \left\{\begin{array}{ll}
\scriptstyle{\sum \limits_{\substack{\scriptstyle{r=} \\ \scriptstyle{-2}}}^{2} \sum \limits_{\substack{\scriptstyle{t=} \\ \scriptstyle{-2}}}^{2} rtF_{2d}(S_i^{(r)},S_i^{(r+1)},S_j^{(t)}, S_j^{(t+1)},\rho)
- \mu_{M_{i}}\mu_{M_{j}}}
& \mkern9mu \rho \neq 1
\\\\
\begin{split}
&	 \scriptstyle{{\sum \limits_{\substack{\scriptstyle{r=} \\ \scriptstyle{-2}}}^{2}}{\sum \limits_{\substack{\scriptstyle{t=} \\ \scriptstyle{-2}}}^{2}} rtF\bigl(\max(S_i^{(r)},S_j^{(t)}),\min(S_i^{(r+1)},S_j^{(t+1)})\bigr)-\mu_{M_{i}}\mu_{M_{j}}}
\end{split}& \mkern9mu \rho = 1
\end{array} \right. \end{split} 
\end{equation*}

%
%
%

\section{Binary Similarity Measures }
These types of similarity measures can only be used for binary alarm sequences. There are many different indices to measure the similarity between two binary sequences \citep{Yang}. In \cite{ci10}, 76 different indices are reported. One issue that needs to be decided beforehand is whether `0's are as important as 1's in similarity. Some methods suggest that they both should be given the same importance, but others only emphasize on 1's. In the context of alarm data, a 0 in the binary sequence means no alarm, which does not provide the operator with any new information. So, for alarm data, indices that are only based on matching 1's  are preferred. An index in this category is the Jaccard index \citep{ci9}.
\subsection{Definition of the Jaccard Index}
For two binary sequences $s_{i}$ and $s_{j}$, the Jaccard similarity index is calculated as  
\begin{equation}
S_{jac}(s_{i}, s_{j})=\frac{a_{i,j}}{a_{i,j}+b_{i,j}+c_{i,j}}
\end{equation}
Here, $a_{i,j}$ is the number of matching 1's (i.e., the total number of times where both $s_i$ and $s_j$ are 1), $b_{i,j}$ is the number of times where $s_i$ is 1 and $s_j$ is 0, and $c_{i,j}$ is the number of times where $s_i$ is 0 and $s_j$ is 1 (hence, $b_{i,j}+c_{i,j}$ represent the total number of mismatches). The Jaccard similarity index is a number in  $[0, 1]$.
\subsection{Analytical Derivations}
To calculate the Jaccard similarity index between two binary alarm sequences, we can use the equivalent concept of probabilistic interpolation of sets of instances.  The Jaccard similarity index can then be rewritten as 
\begin{equation}
\begin{split}
S_{jac}(s_{i}, s_{j})&= \frac{\Pr(\{X_{i}>h_{i}\} \cap  \{X_{j}>h_{j}\})}{\Pr(\{X_{i}>h_{i}\} \cup  \{X_{j}>h_{j}\})}\\
&=\frac{\Pr(X_{i}>h_{i} , X_{j}>h_{j})}{1- \Pr(X_{i}<h_{i} , X_{j}<h_{j})}
\end{split}
\end{equation}
Therefore, the analytical value of the Jaccard similarity index is
\begin{equation*}\label{Pearson_anal_c}
S_{jac}(s_{i}, s_{j})= \left\{\begin{array}{ll}
\frac{F_{2d}(\frac{h_{i}-\mu _{i}}{\sigma _{i}},+\infty,\frac{h_{j}-\mu _{j}}{\sigma _{j}},+\infty,\rho)}{1-F_{2d}(-\infty,\frac{h_{i}-\mu _{i}}{\sigma _{i}},-\infty,\frac{h_{j}-\mu _{j}}{\sigma _{j}},\rho)}
	&  ~\rho \neq 1 \\ \\
\frac{F(\frac{\mathrm{max}(h_{i},h_{j})-\mu_{i}}{\sigma_{i}},+\infty)}{F(\frac{\mathrm{min}(h_{i},h_{j})-\mu_{i}}{\sigma_{i}},+\infty)}
	& ~\rho = 1
\end{array} \right.
\end{equation*}

\begin{figure}
    \begin{minipage}[t]{.24\textwidth}
        \centering
        \includegraphics[width=1.1\textwidth]{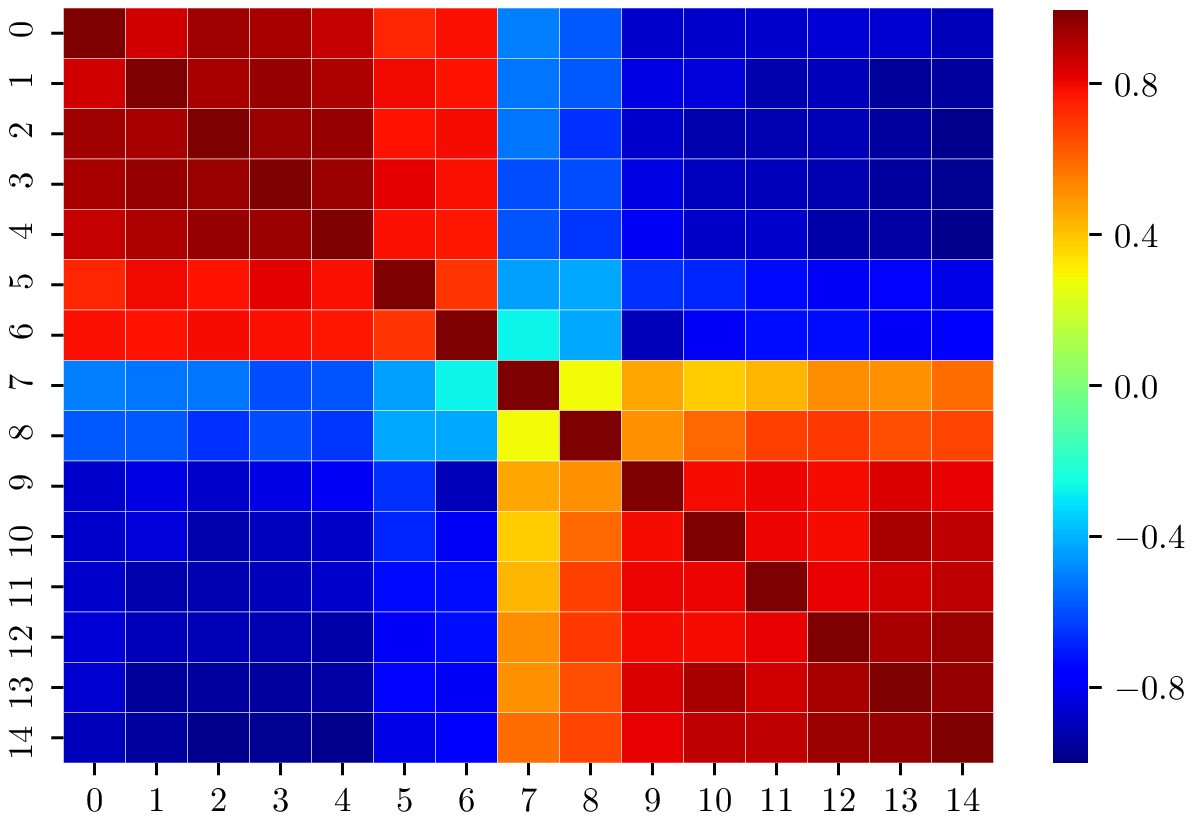}
        \subcaption{Analytic}\label{fig:p1}
    \end{minipage}
    \hfill
    \begin{minipage}[t]{.24\textwidth}
        \centering
        \includegraphics[width=1.1\textwidth]{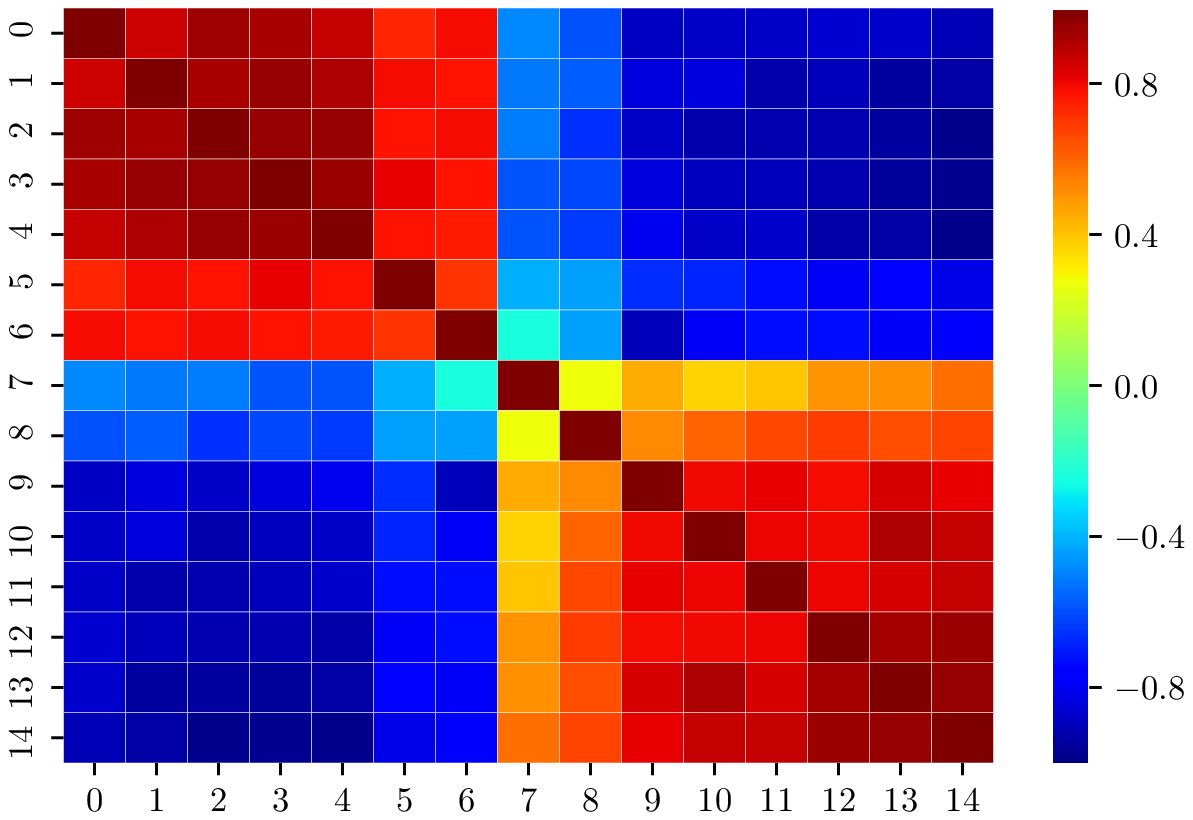}
        \subcaption{Simulation}\label{fig:p2}
    \end{minipage}  
    \caption{The CCM of the original process variables}
    \label{process-var}
    
            \begin{minipage}[t]{.24\textwidth}
        \centering
        \includegraphics[width=1.1\textwidth]{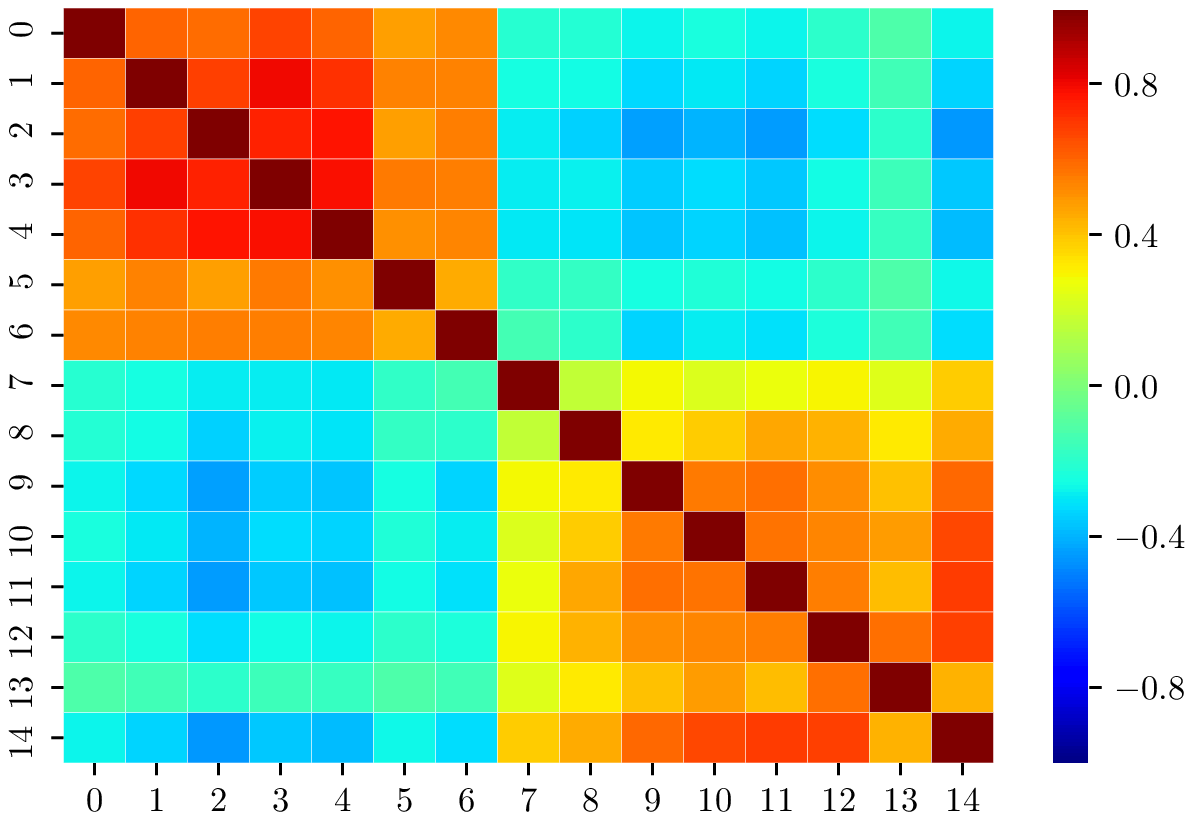}
        \subcaption{Analytic}\label{fig:b1}
    \end{minipage}
    \hfill
    \begin{minipage}[t]{.24\textwidth}
        \centering
        \includegraphics[width=1.1\textwidth]{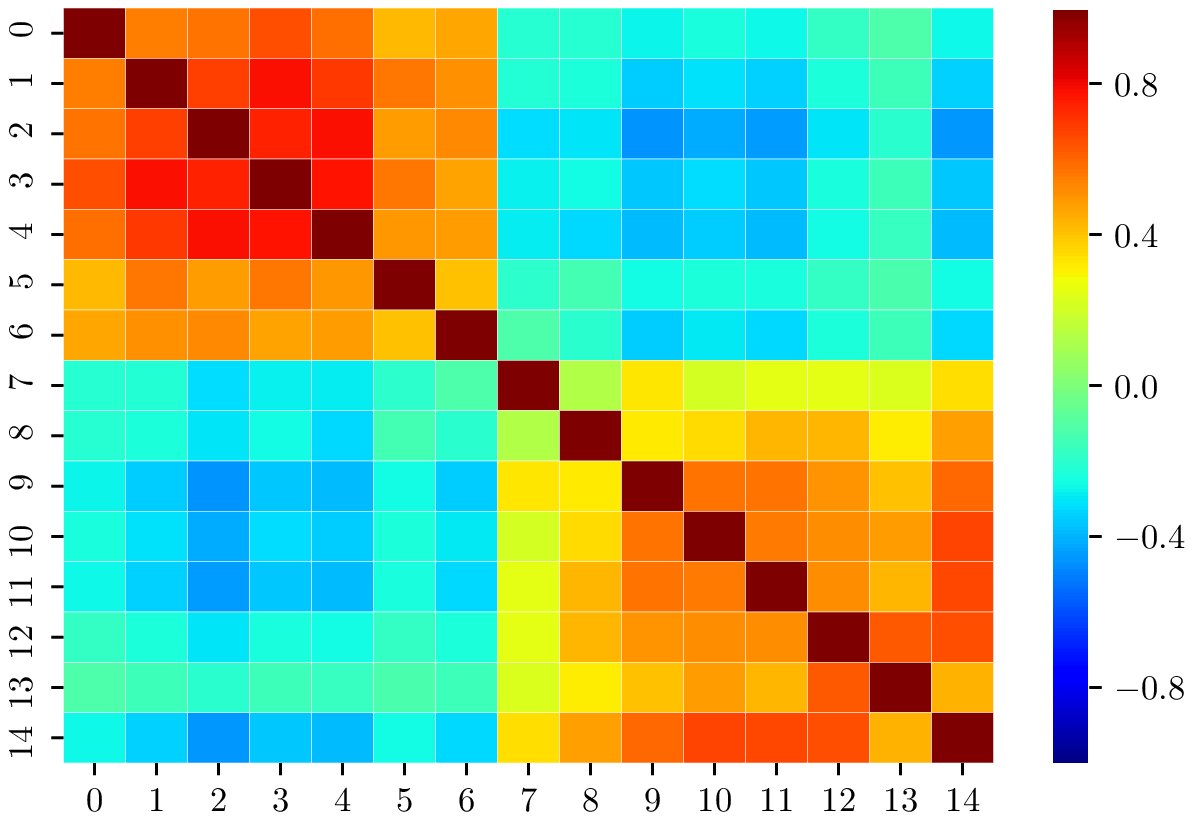}
        \subcaption{Simulation}\label{fig:b2}
    \end{minipage}  
    \caption{The CCM obtained from binary alarm sequences}
    \label{binary-dis}

    \begin{minipage}[t]{.24\textwidth}
        \centering
        \includegraphics[width=1.1\textwidth]{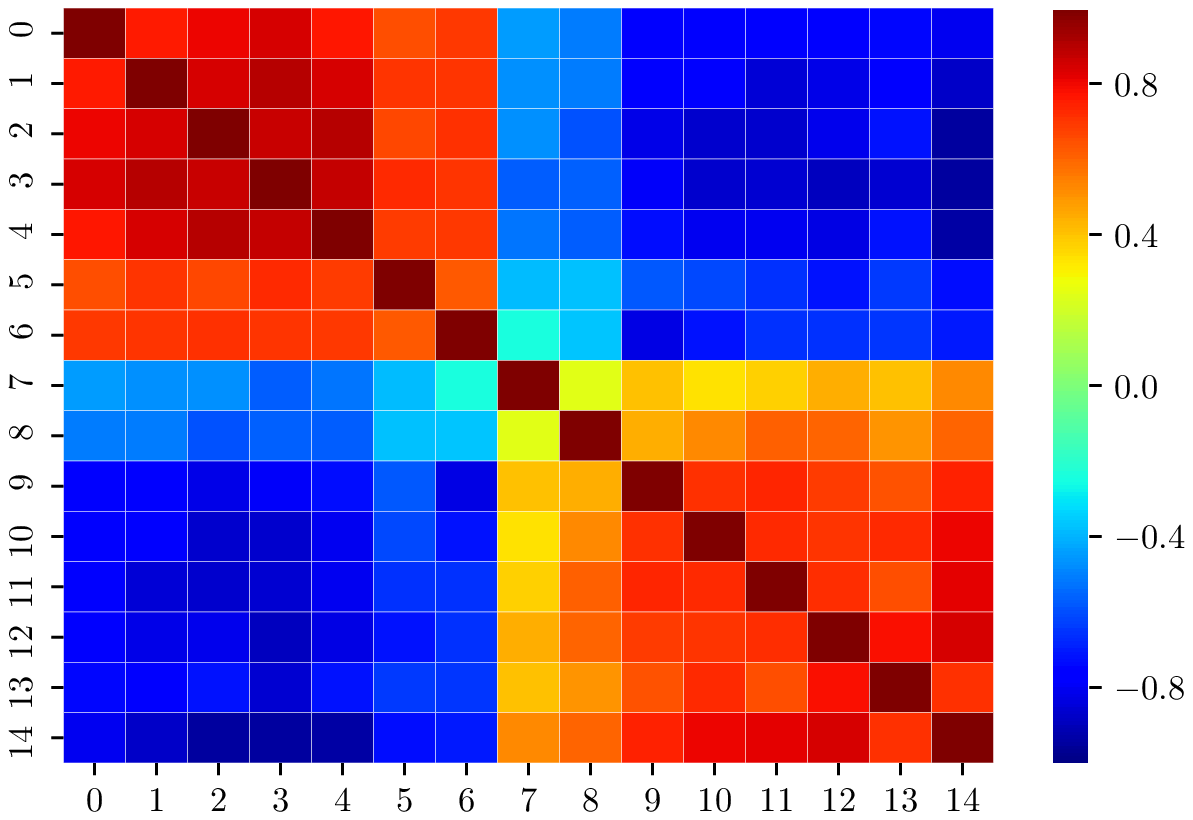}
        \subcaption{Analytic}\label{fig:c1}
    \end{minipage}
    \hfill
    \begin{minipage}[t]{.24\textwidth}
        \centering
        \includegraphics[width=1.1\textwidth]{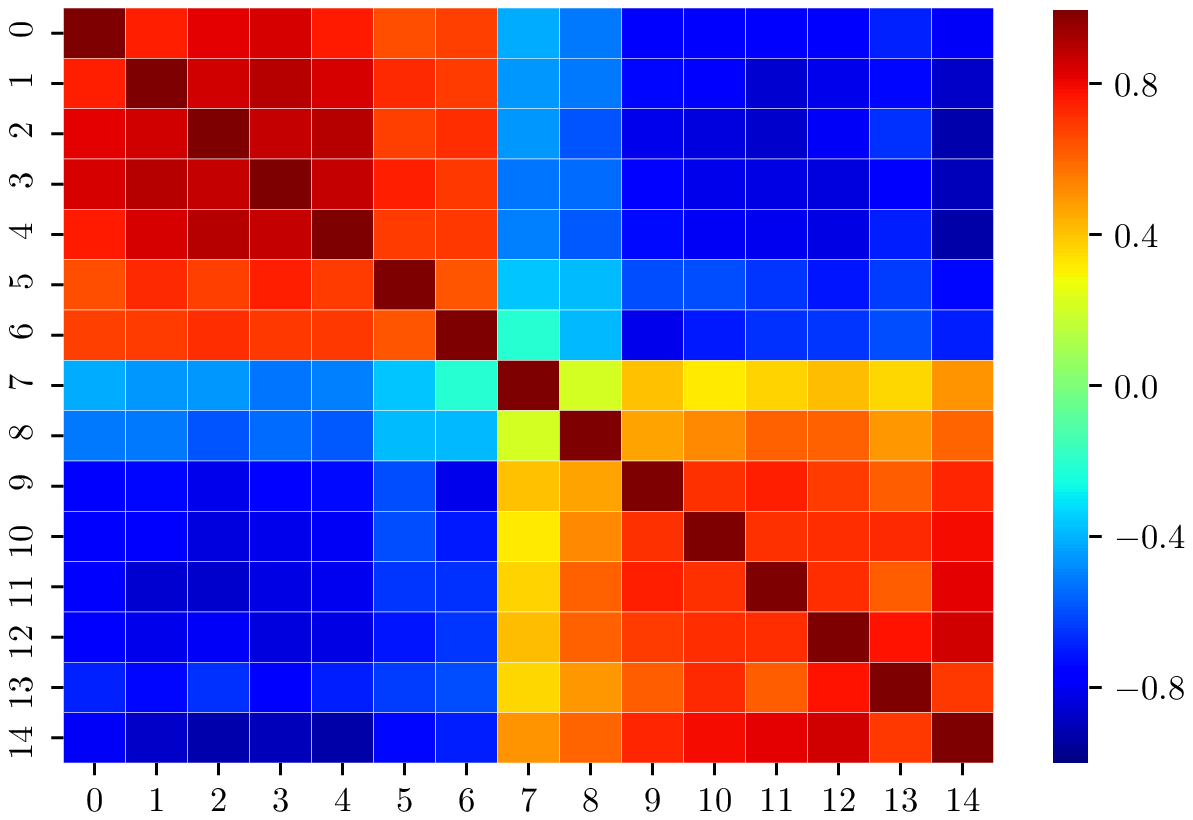}
        \subcaption{Simulation}\label{fig:c2}
    \end{minipage}  
    \caption{The CCM obtained from multivalued alarm sequences}
    \label{multi-valued-dis}
    
    \begin{minipage}[t]{.24\textwidth}
        \centering
        \includegraphics[width=1.1\textwidth]{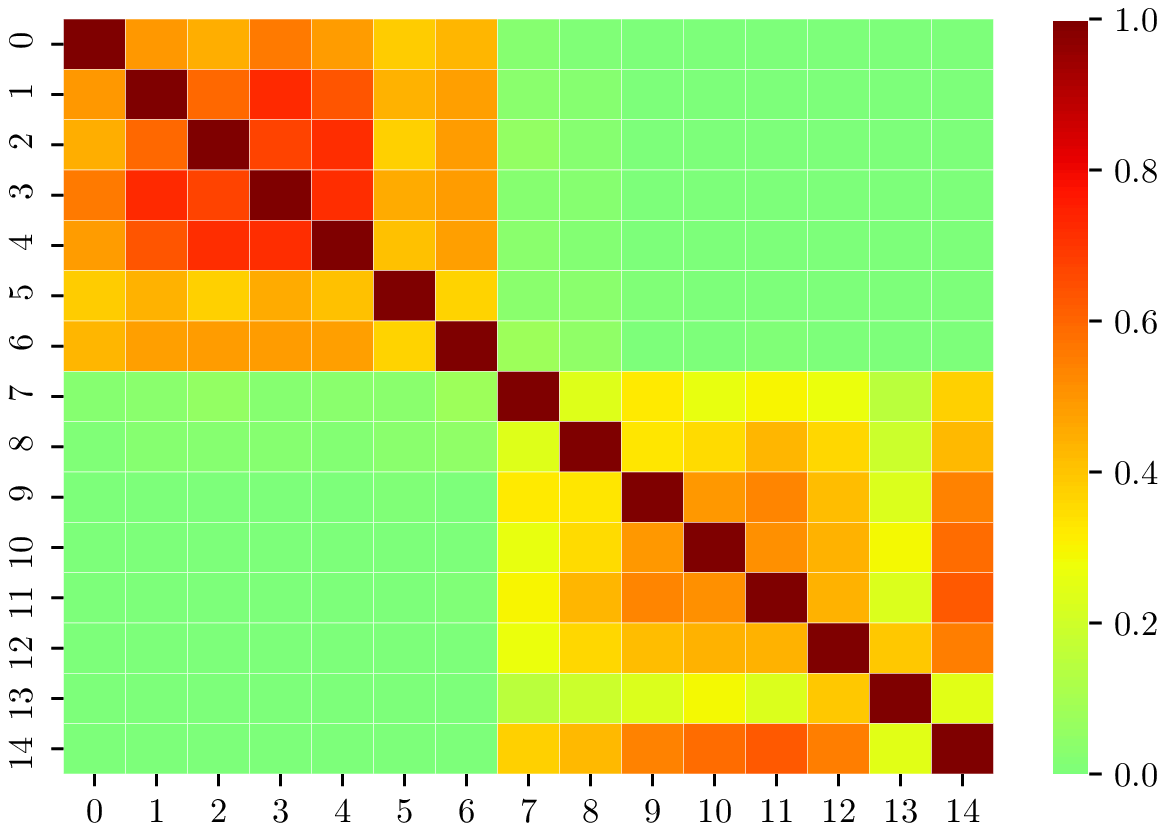}
        \subcaption{Analytic}\label{fig:m1}
    \end{minipage}
    \hfill
    \begin{minipage}[t]{.24\textwidth}
        \centering
        \includegraphics[width=1.1\textwidth]{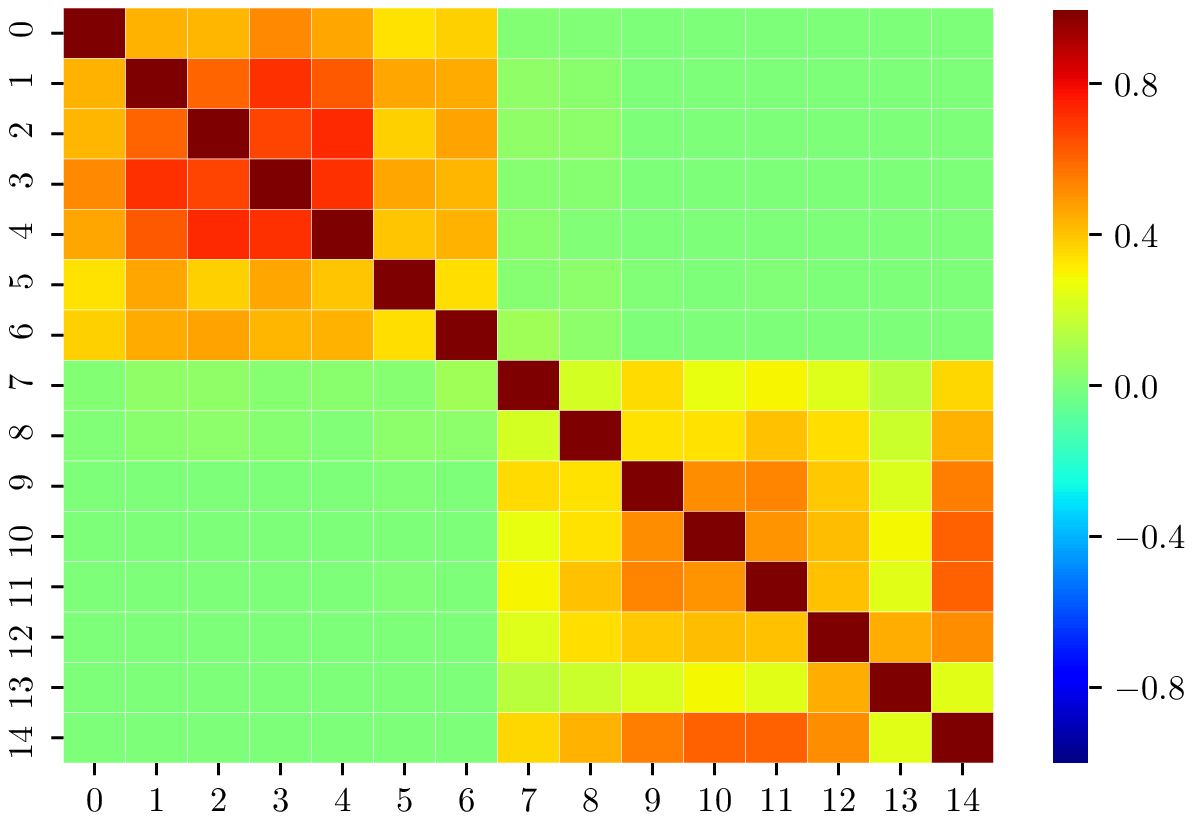}
        \subcaption{Simulation}\label{fig:m2}
    \end{minipage}  
    \caption{The Jaccard similarity color map obtained from binary alarm sequences}
    \label{jaccard}
\end{figure}

\section{Monte-Carlo simulation}
In the previous sections, different measures of similarity between two alarm sequences were discussed. To study how different models of alarm data (binary and multivalued) and various similarity measures are compared, a Monte-Carlo simulation is carried out. For the case study, 15 different correlated PVs with Gaussian distribution but different parameters (mean and variance) are generated~\footnote{The source code of our implementation is available at: \\ \url{https://github.com/kargaranamir/Alarm-Similarity}}.
Also, different values for the pair-wise correlation of these random variables are selected. Fig.~\ref{process-var} shows the correlation color map (CCM) of the original process variables. for validation of the obtained analytical results, the CCM based on analytical calculations (which here is only the correlation initially selected) and that obtained from simulation are depicted. 

To compare different models of alarm data, high alarm limits are considered for all these variables, and their corresponding binary alarm sequences are obtained. The Pearson correlation coefficient, and the Jaccard similarity index for these 15 binary alarm sequences are then calculated. Then, for each PV, 4 alarm limits (HH, HI, LO, LL) are considered and the corresponding multivalued alarm sequence is acquired. Subsequently, the Pearson Correlation coefficient of these multivalued alarm sequences is calculated. The results for each case (both analytical and Monte-Carlo simulation) are reported graphically using the CCM: the Pearson correlation coefficient for binary alarm sequences in Fig.~\ref{binary-dis};  the Pearson correlation coefficient for multivalued alarm sequences in Fig.~\ref{multi-valued-dis}; and the Jaccard similarity index for binary alarm sequences in Fig.~\ref{jaccard}.  Since there are 15 variables, the CCM is a $15 \times 15$ matrix. 

It can be observed that for each case the analytical and simulation-based CCMs are identical. This validates the formulations obtained for different similarity indices.

Moreover, it is expected that the similarity of two alarm variables should, more or less, be proportional to the correlation of the two corresponding PVs. So, it is expected that the CCMs based on alarm variables look similar to that of PVs. This can also be observed from the figures, albeit with different levels of likeness. As it can be seen, the CCM obtained from multivalued alarm sequences is the most similar to the one obtained from the PVs. The other two (based on binary alarm sequences) are not as similar to the CCM of PVs, with the one obtained from the Pearson correlation coefficient more similar than that of the Jaccard similarity index. The reason is that multivalued alarm sequences are richer and carry more information compared to the other two alarm sequences. The mere fact that different alarm types (LL, LO, HI, HH) are treated differently in this sequence, significantly improves the correlations. 

\section{Discussion and Conclusions}
%
In this paper, we studied and compared different methods of similarity analysis of alarms. The differences among the methods are in two aspects. One is how the alarm sequences are constructed. The two common models discussed are binary and multivalued alarm sequences, obtained from the original alarm messages. The second difference is in the method used for calculating similarity. The two widely adopted techniques are the Jaccard similarity index (for binary sequences only), and the Pearson correlation coefficient (for all sequences). For each case, the similarity measure is analytically calculated and verified through Monte-Carlo simulation. 

The analytical results and Monte-Carlo simulations confirm the intuitive result that if two variables are correlated, their alarms are correlated too. Moreover, alarm sequences can be created based on individual alarm tags or combined alarm tags of one PV (e.g., for multivalued cases). The latter carries more information and is expected to yield a better correlation. This is confirmed by simulation too. This means that if it is intended for the correlation of alarms to be similar to the correlation of the underlying PVs, then the Pearson correlation coefficient of multivalued alarm sequences is the best choice. 

The proposed analytical study on alarm similarity measures can be further used to answer several questions regarding the analysis and design of alarm systems. Some of the topics for future studies include:
\begin{itemize}
\item If two PVs are highly correlated, then their corresponding alarms are highly correlated as well. In this case, is it meaningful to configure alarms on both of them? Notice that, one of the deficiencies in alarm systems is related (i.e., correlated) alarms, as they are generally regarded as a nuisance to the operator. If the alarms are to be configured on both PVs, how can one design the alarm limits to make the alarm correlation smaller?

\item If two or more alarms are configured for one single PV (e.g., HH, HI, LO, LL) how should the alarm limits be selected to make the alarms more informative? If, for instance, HH and HI limits are closely selected, then the HH and HI alarms are highly correlated, which again renders them related and not acceptable. 

\item How different alarm configuration techniques, such as delay-timers or filters, impact alarm correlation?
\end{itemize}

\end{document}